\documentclass[a4paper]{jpconf}
\usepackage{graphicx}

\newcommand{\be}{\begin{equation}}
\newcommand{\ee}{\end{equation}}
\newcommand{\ba}{\begin{eqnarray}}
\newcommand{\ea}{\end{eqnarray}}
\newcommand{\pa}{\partial}

\newcommand{\n}[1]{\label{#1}}
\newcommand{\eq}[1]{Eq.(\ref{#1})}

\newcommand{\hh}{\, ,\hspace{0.5cm}}

\begin{document}
\title{Applications of hidden symmetries to black hole physics}

\author{Valeri Frolov}

\address{Institute of Theoretical Physics, Department of Physics
University of Alberta, Edmonton, Alberta, T6G 2G7, CANADA}

\ead{vfrolov@ualberta.ca}

\begin{abstract}  This work is a brief review of applications of
hidden symmetries to black hole physics. Symmetry is one of the most
important concepts of the science. In physics and mathematics the
symmetry allows one to simplify a problem, and often to make it
solvable. According to the Noether theorem symmetries are responsible
for conservation laws. Besides evident (explicit) spacetime
symmetries, responsible for conservation of energy, momentum, and
angular momentum of a system, there  also exist what is called hidden
symmetries, which are connected with higher order in momentum
integrals of motion. A remarkable fact is that black holes in four
and higher dimensions always possess a set (`tower') of explicit and hidden
symmetries which make the equations of motion of particles and light
completely integrable. The paper gives a general review of the recently
obtained results. The main focus is on understanding why at all
black holes have something (symmetry) to hide. \end{abstract}

\section{Introduction} 

In this paper we discuss higher dimensional rotating black holes and 
their properties. We consider black holes with the spherical topology of
the horizon in a spacetime which is either asymptotically flat
(vacuum) or (A)deSitter  (with cosmological constant). We demonstrate
that such solutions of the Einstein equations always possess a
"tower" of symmetries, which make the equations of geodesic motion
completely integrable. This 'tower' is generated by a single object,
which we call a {\em principal conformal Killing-Yano tensor}. This
object is responcible for a set of Killing vectors reflecting a
spacetime symmerty. It also generates a set of Killing tensors,
connected with hidden symmetries of the spacetime. The corresponding
conserved quantities, which are first and second order in the
momentum, form a complete set of integrals of motion which make the
geodesic equation completely integrable. The purpose of the paper is
to describe a general scheme of this construction, and to discuss its
application to concrete problems connected with particle motion and 
field propagation in the higher dimensional black holes. We also
describe a class of metrics which admit the principal conformal
Killing-Yano tensor and its generalizations.

\section{Complete integrability} 

\subsection{Phase space}

Let us first remind three related notions which play an important
role in study of dynamical systems: (i) complete integrability,
(ii) separation of variables, and (iii) hidden symmetries (for more
details see \cite{Arnold,Int}).

A {\em phase space} is a set of three items $\{ M^{2m},\Omega, H\}$.
$M^{2m}$ is a $2m-$dimensional differential manifold. A symplectic
form $\Omega$ is a closed, $d\Omega=0$, non-degenerate 2-form. The
non-degeneracy means that the corresponding matrix of its coefficient
has the rank  $2m$. Locally $\Omega=0$ can be presented in the form
\be
\Omega=d\alpha\, ,
\ee
where $\alpha$ is a 1-form. $H$ is a scalar function on $M^{2m}$
called a {\em Hamiltonian}. We denote coordinates on $M^{2m}$ by
$z^{A}$\  $(A=1,\ldots,2m)$. A set of coordinates which covers all the
manifold is called an {\em atlas}. Figure~\ref{F1} illustrates
these definitions.

\begin{figure}
\begin{center}
\includegraphics[width=8cm]{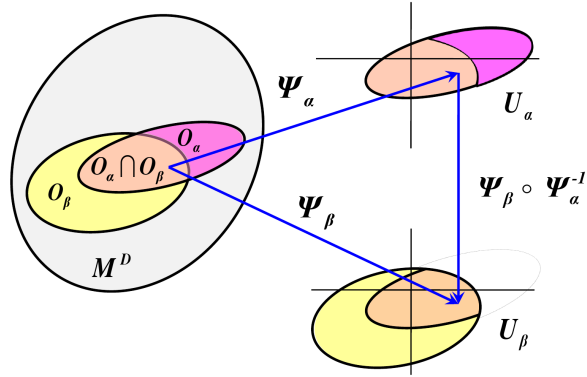}
\end{center}
\caption{\n{F1}
Coordinates on $n-$dimensional manifold are maps
$\Psi_{\alpha}$ of open regions $O_{\alpha}$ on the regions
$U_{\alpha}$ in the $n-$dimensional Euclidean space $R^n$. The map
$\Psi_{\beta}\circ \Psi^{-1}_{\beta}$ determines a transformation
between two coordinate systems. 
} 
\end{figure}

Since the symplectic form is non-degenerate, the tensor $\Omega_{AB}$
has an inverse one $\Omega^{AB}$, defined by the relation
$\Omega^{AB}\Omega_{BC}=\delta^A_C$. The tensors $\Omega_{AB}$ and
$\Omega^{AB}$ can be used to relate tensors with upper and lower
position of indices. For example, 
\be
\eta^A=\Omega^{AB}H_{,B}\, 
\ee
is a vector generating the Hamiltonian flow. The equation of motion of
a particle in the phase space is
\be\n{he}
\dot{z}^A=\eta^A\, .
\ee
{Poisson bracket} for two functions on the phase space $A$ and
$B$ is defined as
\be
\{A,B\}=\Omega^{AB}A_{,A}B_{,B}\, .
\ee
One says that these function are in {\it involution} if their Poisson
bracket vanishes.
The Poisson brackets for any three functions $A$, $B$, and $C$ on the
symplectic manifold obey the {\it Jacobi identity}
\be\n{JID}
\{\{A,B\},C\}+\{\{B,C\},A\}+\{\{C,A\},B\}=0\, .
\ee

If $F(z)$ is a function on the phase space, the equation of motion
\eq{he} determines its evolution $F_t=F(z(t))$
\be
\dot{F}_t=\{H,F_t\}\, .
\ee
A function $F$ for which $\{H,F\}=0$ is an integral of motion of the
Hamiltonian system. The Hamiltonian itself is a trivial integral of motion.

According to {\em Darboux theorem}, in the vicinity of any point of the
phase space it is always possible to choose {\em canonical
coordinates} $z^{A}=(q_1,q_2,\ldots, q_m,p_1,p_2,\ldots, p_m)$ in
which the symplectic form is
\be
\Omega=\sum_{i=1}^m dp_i\wedge dq_i\, .
\ee
In such canonical coordinates many relations are simplified and take
a well known form. A set of canonical coordines which covers the phase
space is called a {\em canonical atlas}. 

\subsection{Integrability}

Integrability of equations of a dynamical system generically means
that these equations can be solved by quadratures. Integrability is
linked to `existence of constants of motion'. For the integrability it
is important to know: (i) How many constants of motion exist? (ii) How
precisely are they  related? and (iii) How the phase space is foliated
by their level sets? 

A system of differential equations is said to be integrable by
quadratures if its solution can be found after a finite number of
steps involving algebraic operations and integration. 
The following theorem (Bour, 1855; Liouville, 1855) establishes
conditions required for the complete integrability of a
Hamiltoniam system:
\newline
{\it If a Hamiltonian system with $m$ degrees of freedom has $m$
integrals of motion $F_1=H,F_2,\ldots, F_m$ in involution which are
functionally independent on the intersection sets of the $m$
functions, $F_i=f_i$, the solutions of the corresponding Hamiltonian
equations can be found by quadratures.}

The main idea behind Liouville's theorem is that the first integrals
of motion $F_i$ can be used as $m$ coordinates. The involution
condition implies that the $m$ vector fields generated by gradients of
$F_i$ commute with each other and provide a choice of canonical
coordinates. In these coordinates, the Hamiltonian is effectively
reduced to a sum of $m$ decoupled Hamiltonians that can be integrated.

\begin{figure}
\begin{center}
\includegraphics[width=8cm]{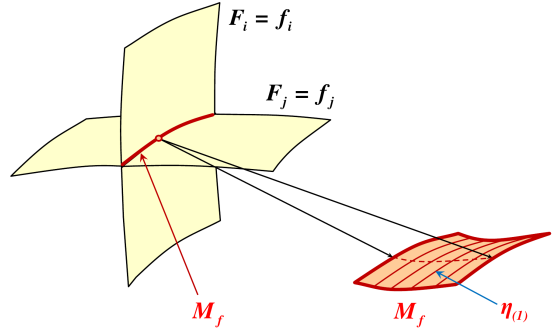}
\end{center}
\caption{\label{F2}Illustration to the proof of Liouville theorem.} 
\end{figure}

We denote by $M_f$ an intersection of the level sets $F_i=f_i$ for
$m$ integrals of motion (see Figure~\ref{F2}). Since the integrals of
motion are independent, the tangent  to $M_f$ vectors 
$X_i^A=\Omega^{AB}F_{i,B}$ are linearly independent and  $M_f$ is a
$m-$dimensional submanifold of the phase space.  The condition that
the integrals of motion are in involution implies
\be
[X_i,X_j]=0\, .
\ee
One also has
\be\n{oxx}
\Omega_{AB}X_i^A X_j^B=\{ F_i,F_j\}=0\, .
\ee
Let us use the equations $F_i(p,q)=f_i$ to get $p_i=p_i(f,q)$. 
Denote
\be
S(F,q)=\int_{q^0}^{q}\sum _{i=1}^m p_i(f,q) dq_i\, .
\ee
The integral is taken over a path on $M_f$ connecting two its points
$q^0$ and $q$. The relation \eq{oxx} implies that $S(F,q)$ does not
depend on the choice of the path. Denote
\be
\Psi_i={\pa S\over \pa F_i}\, ,
\ee
then
\be
dS=\sum_{i=1}^m\left( \Psi_i dF_i+p_i dq_i\right)\, .
\ee
Since the symplectic form $\Omega$ is closed, one has $d^2S=0$ and
hence
\be
\Omega=\sum_{i=1}^m  dp_i \wedge dq_i=\sum_{i=1}^m  d\Psi_i \wedge
dF_i\, .
\ee
This result means that there exists a canonical transformation
$(p_i,q_i)$ to $(F_i,\Psi_i)$. To obtain `new' canonical coordinates
two operations are required: (1) finding of $p_i=p_i(f,q)$, and (2) 
calculation of some integrals. In the new variables the Hamiltonian
equations take the form
\ba
\dot{F}_i&=&\{ H,F_i\}=0\, ,\\
\dot{\Psi}_i&=&\{ H,\Psi_i\}={\pa H\over \pa F_i}=\delta_i^1\, .
\ea
The solution of this system is trivial
\be
F_i=\mbox{const}\hh \Psi_i=a_i+\delta_i^1 t\, .
\ee

Complete integrability and chaotic motion are at the two ends of
`properties' of a dynamical system. The integrability is exceptional,
while the chaoticity is generic. In all cases, integrability seems to
be deeply related to some symmetry, which might be partially hidden.
The existence of integrals of motion reflects the symmetry.

Important known examples of completely integrable mechanical systems
include:
\begin{enumerate}
\item Motion in Euclidean space under a central potential;
\item Motion in the two Newtonian fixed centers;
\item Geodesics on an ellipsoid;
\item Motion of a rigid body about a fixed point;
\item Neumann model\footnote{The Lagrangian of the Neumann model is
\be
L={1\over 2}\sum_{k=1}^N \left[ \dot{x}_k^2-a_k x_k^2
+\Lambda(x_k^2-1)\right]\, . 
\ee
}.
\end{enumerate}

\section{Separation of variables and integrability}

Complete integrability of the Hamiltonian systems is closely related
with the separation of variables in the Hamilton-Jacobi equation.
For a Hamiltonian $H(P,Q)$, \ $P=p_1,\ldots,p_m$ and
$Q=q_1,\ldots,q_m$, the {\it Hamilton-Jacobi equation} is
\be\n{SH}
H(\pa_{Q}S,Q)=0\hh \pa_{Q}S=(\pa_{q_1}S,\ldots,\pa_{q_m}S)\, .
\ee
If a variable $q_1$ and a derivative $\pa_{q_1}S$ enter this equation
in a form of single expression $\Phi_1(\pa_{q_1}S,q_1)$, than one says that
the variable $q_1$ is separated. In such a case one may try to search
a solution in the form 
\be\n{S1}
S=S_1(q_1)+S'(q_2,\ldots,q_m)\, .
\ee
Putting
\be\n{SS1}
\Phi_1(\pa_{q_1}S,q_1)=C_1\, ,
\ee
one obtains an equation with a less
number of variables
\be\n{SH1}
H_1(\pa_{q_2}S',\ldots,\pa_{q_m}S',q_2,\ldots,q_m;C_1)=0\, .
\ee
Let $S'(q_2,\ldots,q_m;C_1)$ be a solution of this equation depending
on a parameter $C_1$, then \eq{S1} is a solution of \eq{SH} when $S_1$
obeys an ordinary differential equation \eq{SS1}, which is easily
solvable by quadratures.
If a variable $q_2$ can be separated in a new equation \eq{SH1} one
can repeat this procedure again. One says that the
Hamilton-Jacobi equation \eq{SH} allows a {\it complete separation of
variables} if after $m$ steps we obtain a solution of the initial
equation \eq{SH} which contains $m$ constants $C_i$
\be
S=S_1(q_1,C_1)+S_2(q_2;C_1,C_2)+\ldots +S_m(q_m;C_1,\ldots,C_m)\, .
\ee
In this case one obtains a complete solution of the
Hamilton-Jacobi equation which depends on $m$ parameters and the
corresponding Hamilton equations are integrable by quadratures
({\it Jacobi theorem}).

The constants $C_1,\ldots,C_m$ for a completely separable
Hamilton-Jacobi equation can be considered as functions on the phase
space where they are integrals of motion. In a case, when these
integrals on motion are independent and in involution, the system is
completely integrable in the sense of Liouville.

\section{Particle motion in General Relativity}

\subsection{Equation of motion in the Hamiltonian form}

Consider a particle motion of mass $m$ in the gravitational field. 
Its equation of motion is
\be\n{eqm}
m\,\frac{D^2 x^{\mu}}{d\tau^2}= 0\, .
\end{equation}
Here, ${D/d\tau}$ is the covariant derivative with respect
to the proper time ${\tau}$.
Introduce the affine parameter $\lambda=\tau/\mu$ and denote a
derivative with respect to it by a dot. The Lagrangian for \eq{eqm} is
\be
L={1\over 2}g_{\mu\nu}\dot{x}^{\mu}\dot{x}^{\nu}\, .
\ee
The momentum $p_{\mu}$ is defined as
follows
\be
p_{\mu}={\pa L\over \pa\dot{x}^{\mu}}=g_{\mu\nu}\dot{x}^{\nu}\, .
\ee

Denote by $D$ the dimensionality of the spacetime. 
Coordinates $(p_{\mu},x^{\mu})$ are canonical on the phase space
$M^{2D}$. In these coordinates the symplectic form is
\be
\Omega=\sum_{\mu=1}^D dp_{\mu}\wedge dx^{\mu}\, .
\ee
The Hamiltonian of the particle is
\be
H={1\over 2}g^{\mu\nu}p_{\mu}p_{\nu}\, .
\ee
It gives trivial integral of motion
\be
H={1\over 2}m^2\, .
\ee
For null rays one must put $m=0$.
The Hamiltonian equations of motion
\ba
\dot{x}^{\mu}&=&\{H,x^{\mu}\}=g^{\mu\nu}p_{\nu}\, ,\\
\dot{p}_{\mu}&=&\{H,p_{\mu}\}=-{1\over 2} g^{\nu\lambda}_{\ \
,\mu}p_{\nu} \, ,
\ea
are equivalent to the geodesic equation \eq{eqm}, which can be written
in the form
\be
p^{\nu}p_{\mu;\nu}=0\, .
\ee

\subsection{Integrals of motion and Killing tensors}

Consider a special monomial in the momentum on the phase space of the
relativistic particle of the form
\be
{\cal K}=K^{\mu_1\ldots \mu_s}p_{\mu_1}\ldots p_{\mu_s}\, .
\ee
A condition that this is an integral of motion implies
\be\n{kte}
K^{(\mu_1\ldots \mu_s;\nu)}=0\, .
\ee
The symmetric tensor of the rank $s$, $K_{\mu_1\ldots \mu_s}$, which
obeys the equation \eq{kte}, is called a {\it Killing tensor}. The
metric $g_{\mu\nu}$ is a trivial example of the Killing tensor of rank
2.

Suppose we have two integrals of motion, ${\cal K}_{(s)}$ and 
${\cal K}_{(t)}$, connected with  the Killing tensors
$K^{\mu_1\ldots\mu_s}_{(s)}$ and $K^{\nu_1\ldots\nu_t}_{(t)}$,
respectively. Using the {\it Jacobi identity} \eq{JID} it is easy to
check that $\{{\cal K}_{(s)},{\cal K}_{(t)}\}$ also commutes with the
Hamiltonian $H$ and hence is an integral of motion. This commutator is
a monomial of the power $s+t-1$ in the momentum. The corresponding
Killing tensor of the rank $s+t-1$ is
\be
[K_{(s)},K_{(t)}]=K_{(s+t-1)}\, ,\n{KKK}
\ee
\be\label{Schouten}
K_{(s+t-1)}^{\mu_1\ldots\mu_{s-1}\lambda \nu_1\ldots\nu_{t-1}}=
s\,K_{(s)}^{\epsilon(\mu_1\ldots\mu_{s-1}}
\partial_{\epsilon}K_{(t)}^{\lambda\nu_1\ldots\nu_{t-1})}
-t\,K_{(t)}^{\epsilon(\nu_1\ldots\nu_{t-1}}
\partial_{\epsilon}K_{(s)}^{\lambda\nu_1\ldots\mu_{s-1})}
\ee
The introduced operation for the Killing tensors is known as the 
{\it Schouten-Nijenhuis brackets}.   Killing tensors form a Lie
subalgebra of a Lie algebra of all totally symmetric contravariant
tensor fields on the manifold with respect to these operations. The
Killing tensors for which the Schouten-Nijenhuis bracket vanishes are
said to be in involution.

In a simplest case  when a monomial is of the first order in the
momentum, the Killing tensor coincides with the Killing vector, and 
the  {\it Schouten-Nijenhuis bracket} reduces to a usual commutator of
two vector fields.

If there exist $D$ non-degenerate functionally independent Killing
tensors\footnote{Some of them can be Killing vectors.}
 in involution then the geodesic equations in $D-$dimensional
spacetime are completely integrable.

Geodesic motion in the gravitational field of the most general
solution describing a rotating black hole in  4 and higher dimensional
spacetimes, which are asymptotically either flat or (A)dS,  is a new
class of physically interesting completely integrable systems.  

\section{Killing-Yano tensors}

\subsection{Definitions}

Killing tensors are natural symmetric generalizations of a Killing
vector. Let us discuss another important generalization, known as a
Killing-Yano tensor. We define first a {\it conformal Killing-Yano
tensor} of rank $p$. It is an antisymmetric tensor $k_{\mu_1\ldots
\mu_p}$ which obeys the following equation
\be\label{CKY}
\nabla_{(\mu_1}k_{\mu_2)\mu_3 \ldots \mu_{p+1}}=\ 
g_{\mu_1 \mu_2}\tilde{k}_{\mu_3 \ldots \mu_{p+1}}-
(p-1)g_{[\mu_3(\mu_1}\tilde{k}_{\mu_2) \mu_4\ldots \mu_{p+1}]}\, .
\end{equation}
By tracing both sides of this equation one obtains 
\be
\tilde{k}_{\mu_2 \mu_3 \ldots \mu_{p}}={1\over
D-p+1}\nabla^{\mu_1}k_{\mu_1 \mu_2
\ldots \mu_p}\, .
\ee
In the case when $\tilde{k}_{\mu_2 \ldots \mu_p}=0$ one has a
{\em  Killing--Yano} (KY) tensor. For the KY tensor ${\bf
k}$ the quantity
\begin{equation}\label{Lprop}
L_{\mu _1 \mu_2 \ldots \mu_{p-1}}=k_{\mu_1 \mu_2 \ldots \mu_p}p^{\mu_p}\,, 
\end{equation}
is parallel-propagated along the geodesic.

The `square' {\bf K} of  KY tensor ${\bf k}$, defined by the relation
\begin{equation}\label{CKT} 
K_{\mu\nu}=({\bf k}\circ {\bf k})_{\mu\nu}\equiv {(p-1)!}\,k_{\mu \mu_2 \ldots
\mu_p}k_{\nu}^{\ \,\mu_2 \ldots\mu_p}\, ,
\end{equation}
is a Killing tensor. Notice that not an arbitrary Killing tensor can be
written in the form \n{CKT}. 

\subsection{Properties of conformal Killing-Yano tensors}

\begin{figure}
\begin{center}
\includegraphics[width=8cm]{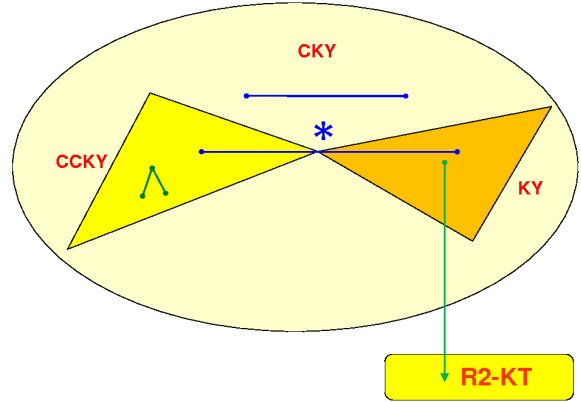}
\end{center}
\caption{\label{F3} Schematical illustration of the properties of
closed conformal Killing-Yano tensors (CCKY), Killing-Yano tensors
(KY), and Killing tensors of rank 2 (R2-KT). 
} 
\end{figure}

The (conformal) Killing-Yano tensors have the following properties:
\begin{enumerate}
\item Hodge dual of a conformal Killing-Yano tensor is a conformal
Killing-Yano tensor;
\item Hodge dual of a closed conformal Killing-Yano tensor is a
Killing-Yano tensor;
\item  External product of two closed conformal Killing-Yano tensor is
a closed conformal Killing-Yano tensor.
\end{enumerate}
Figure~\ref{F3} schematically illustrates these properties. The last
of these statements was proved in \cite{KKPF,F}.

\subsection{Principal conformal Killing-Yano tensor}

Consider an antisymmetric tensor of rank 2 $h_{\mu\nu}$ which obeys the
following equation
\be\n{pcky}
\nabla_{\gamma} h_{\mu\nu}=g_{\gamma\mu}\xi_{\nu}-g_{\gamma\nu}\xi_{\mu}\, .
\ee
This is a conformal Killing-Yano tensor. Equation (\ref{pcky})
implies the following relations
\be\n{pcky1}
\nabla_{[\gamma} h_{\mu\nu]}=0\hh \xi_{\mu}={1\over D}\nabla^{\nu}
h_{\nu\mu}\, .
\ee
The first of these relations means that ${\bf h}$ is closed. Here
$D=2n+\varepsilon$ is the spacetime dimension. For even number of
dimensions $\varepsilon=0$, while for the odd number $\varepsilon=1$.
The tensor ${\bf h}$ is non-degenerate if its matrix rank is $2n$.
A {\it principal conformal Killing-Yano tensor} is a non-degenerate
closed conformal Killing-Yano tensor of rank 2. The existence of the
principal conformal Killing-Yano tensor for the most general known
solution \cite{CLP} for higher dimensional rotating black holes with spherical
topology of the horizon was proved in \cite{FK,KF}.

It is possible to show
that the vector $\xi^{\mu}$ defined by the second equation in
(\ref{pcky1}) is a Killing vector. We call it a {\it primary Killing
vector} \cite{CQG}.

\section{Killing-Yano `tower'}

In a spacetime with a principal conformal Killing-Yano tensor it is
possible to construct a set of Killing vectors and tensors, which we
call a {\it Killing-Yano `tower'}. The idea of this constuction is
following. For a given principal conformal Killing-Yano tensor ${\bf
h}$ one can define the a set  of $n-1$ objects
\be
{\bf h}^{\wedge j}=\underbrace{{\bf h}\wedge {\bf h}\wedge\ldots
\wedge {\bf h}}_{j\mbox{ times}}\hh j=1,\ldots, n-1\, .
\ee
Each of these objects is a closed conformal Killing-Yano tensor, so
that their Hodge dual are Killing-Yano tensors
\be
{\bf k}_{(j)}=*{\bf h}^{\wedge j}\, .
\ee
By taking `squares' of these tensors one obtains $n-1$ Killing tensors
of the second rank
\be
{\bf K}_{(j)}={\bf k}_{(j)}\circ {\bf k}_{(j)}\, .
\ee
If $\xi^{\nu}$ is a primary Killing vector then it is possible to show that 
\be
\xi_{(j)}^{\mu}={K}_{(j)\nu}^{\mu}\xi^{\nu}\, 
\ee
are again Killing vectors. Thus one obtains $n$ Killing vectors. In
the odd dimensional spacetime there exists an additional Killing
vector $\eta=*{\bf h}^{\wedge n}$. A set of constructed
$n+\varepsilon$ Killing vectors, $n-1$ Killing tensors, and one
trivial Killing tensor (${\bf g}$) gives $D$ conserved quantities
for the geodesic motion. It is possible to show that the corresponding
integrals of motion are independent
and in involution. Thus geodesic motion equations are completely
integrable in a spacetime with a principal conformal Killing-Yano
tensor \cite{PKVK,KKPV,o1}.

\section{Canonical form of metric}

\subsection{Canonical coordinates}

In a spacetime with a principal conformal Killing-Yano tensor ${\bf
h}$ there exists a special convenient choice of coordinates. Consider
the following eigen-value problem
\be
h^{\mu}_{\ \nu} m_{\pm a}^{\nu}=\mp i x_a m_{\pm a}^{\mu}\, .
\ee 
Complex eigen-vectors $m_{\pm a}^{\mu}$ can be written as 
\be
m_{\pm a}^{\mu}=e_a^{\mu}\pm i e_{\hat{a}}^{\mu}\, ,
\ee
where ${\bf e}_{a}$ and ${\bf e}_{\hat{a}}$ are mutually orthogonal
normalised real vectors. A non-degerate ${\bf h}$ has $n$ different
eigen-values $x_a$ ($a=1,\ldots,n$), and the corresponding
eigen-space for each of these eigen-values is two dimensional (see
\cite{HOY:09}). One can use $x_a$ as $n$ coordinates on the
spacetime manifold. We call them {\it Darboux coordinates}. For each
of the Killing vectors $\xi$ from the Killing-Yano `tower' one can
introduce a Killing parameter so that the integral line of this
vector field is a solution of the equation
\be
{dx^{\mu}\over d\psi}=\xi^{\mu}\, .
\ee
This gives us $n+\varepsilon$ Killing coordinates $\psi_j$
($j=0,\ldots,n-\varepsilon$). Total number of Darboux and Killing
coordinates is $D$, which is sufficient for using as coordinate system
in the $D$ dimensional spacetime manifold. We call these coordinates
{\it canonical}. 

\subsection{Off-shell canonical metric}

In these coordinates the metric of the spacetime is
of the form \cite{o2,KFK,hoi1}
\be\n{off}
ds^2=\sum_{a=1}^n \left[ {U_a\over X_a}(dx_a)^2+{X_a\over U_a}\left(
\sum_{j=0}^{n-1} A_a^{(j)}d\psi_j\right)^2\right] -{\varepsilon c\over
A^{(n)}}\left(\sum_{j=0}^n A^{(j)}d\psi_j\right)^2\, .
\ee
Here
\ba
&&U_a=\prod_{b\ne a}(x_b^2-x_a^2)\hh
X_a=X_a(x_a)\, ,\\
&&\prod_{a=1}^n (1+\lambda x_a^2)=\sum_{j=0}^n \lambda^j A^{(j)}\hh
(1+\lambda x_b^2)\prod_{a=1}^n (1+\lambda x_a^2)=\sum_{k=0}^{n-1}
 \lambda^k A^{(k)}_b\, .
\ea

A potential ${\bf b}$, which generates the principal conformal
Killing-Yano tensor ${\bf h}$,
\be
{\bf h}=d{\bf b}\, ,
\ee
in the canonical coordinates is
\be
{\bf b}={1\over 2}\sum_{k=0}^{n-1} A^{(k+1)} d\psi_k\, .
\ee

The metric (\ref{off}) is of the algebraical type D. As we indicated
above, geodesic equations in this metric are completely integrable.
Besides this, it also has the following nice properties. In the metric
(\ref{off}):
\begin{enumerate}
\item Hamilton-Jacobi and Klein-Gordon equations allow
complete separation of variables \cite{FKK};
\item Massive Dirac equation is separable \cite{OY};
\item Stationary string equations are completely integrable \cite{SKF};
\item Tensorial gravitational perturbation equations are separable
\cite{oy2};
\item Equations of the parallel transport along timelike and null
geodesics can be integrated \cite{CFK,KFC};
\item Equations for charged particle motion in such a spacetime in the
presence of a test electromagnetic field are completely integrable
provided this field is generated by the primary Killing vector
\cite{FFKK}.
\end{enumerate}

\subsection{On-shell metric}

The metric (\ref{off}) is valid for any geometry with a principal
conformal Killing-Yano tensor. We call this metric {\it off-shell},
since it does not obey the Einstein equations. Let us consider now
on-shell geometry, that is assume that the metric obeys the equations
\be\n{eeq}
R_{\mu\nu}=(D-1)\Lambda g_{\mu\nu}\, .
\ee
$\Lambda$ is the cosmological constant. The Einstein equations
restrict arbitrary functions $X_a(x_a)$, which enter (\ref{off}), so
that they take the form of polynomials \cite{CLP,HHOY}
\be
X_a=b_a x_a+\sum_{k=0}^n c_k x_a^{2k}\, .
\ee
As a result, the solution depends on $D-\varepsilon$ arbitrary
parameters. This solution coincides with the most general solution for
higher dimensional black holes in either asymptotically flat
($\Lambda=0$), or asymptotically (A)dS spacetime, obtained in
\cite{CLP}. Arbiraty constants, which enter this solutions are: the
cosmological constant $\Lambda$, the mass $M$, $(n-1+\varepsilon)$
rotation parameters, and $(n-1-\varepsilon)$ `NUT' parameters. In the
4D case this is a Kerr-NUT-(A)dS metric.

\subsection{Further developments}

Let us mention two recent developments of the above described results.

(1) In our discussion we assumed that the closed conformal Killing-Yano
tensor is non-degenerate. In particular, this implies that there
exists a set of $n$ different eigen-values of ${\bf h}$, which we used
as Darboux coordinates. In a degenerate case, there may exist
several eigen-values which are constants, and some of these constants
can vanish. The general form of the canonical metric for such
degenerate cases was constucted in \cite{HOY:09}.

(2) We discussed vacuum (with cosmological constant) solutions of the
higher dimensional Einstein equations. An interesting generalization
to a non-vacuum case was obtained recently in \cite{KKY:09}. The
authors consider a {\it five dimensional} minimally coupled gauged
supergravity, which includes gravity and the Maxwell field with a
Chern-Simons term. The corresponding Largrangian density is
\be
L=*(R+\Lambda)-{1\over 2} F\wedge *F+{1\over 3\sqrt{3}}F\wedge F\wedge
A\, .
\ee
The corresponding Einstein-Maxwell equations are
\ba
&&R_{\mu\nu}+{1\over 3}\Lambda={1\over 2}(F_{\mu\lambda}F_{\nu}^{\
\lambda}-{1\over 6}g_{\mu\nu} F^2)\, ,\\
&& dF=0\hh d*F-{1\over \sqrt{3}}F\wedge F=0\, .
\ea
The main result of this work is the following. One can modify the
covariant derivative by including a non-vanishing torsion $T={1\over
\sqrt{3}}*F$, and generalize the equation (\ref{pcky}), by subsituting
the modified derivative insted of the covariant one. The authors
demonstrated that the generalized principal conformal Killing-Yano
tensor generates a `tower' of integrals of motion which provides
complete integrability of a charged particle motion in these spaces.
An interesting example of a charged rotating black hole solution in
this theory was obtained in \cite{CCLP}. It is interesting, that
the corresponding metric is of a general algebraical type.

\subsection{Acknowledgments}
The author is  grateful to the Natural Sciences and
Engineering Research Council of Canada and the Killam Trust for
their support.


\medskip
\section*{References}
\begin{thebibliography}{9}

\bibitem{Arnold} Arnold V I 1989 {\it Mathematical Methods of
Classical Mechanics} Springer

\bibitem{Int} Babelon O, Bernard D, and Talon M 2003 {Introduction to
Classical Integrable Systems} Cambridge Univ. Press.


\bibitem{KKPF} Krtou\v s P, Kubiz\v n\'ak D, Page D N, and Frolov V P 2007
{\it J.~High Energy Phys.} {\bf 02}  004

\bibitem{F}Frolov V P 2008 {\it Prog.Theor.Phys.Suppl.} {\bf 172} 210


\bibitem{CLP} Chen W, L\"u H, and Pope C N  2006
{\it Class. Quant. Grav.} {\bf 23} 5323 

\bibitem{FK} Frolov V P and Kubiz\v n\'ak D  2007 {\it Phys.Rev.Lett} 
{\bf 98} 011101 


\bibitem{KF} Kubiz\v n\'ak D and Frolov V P 2007 {\it Class. Quant.
Grav.} {\bf 24} F1


\bibitem{CQG} Frolov V P and Kubiz\v n\'ak D 2008  
{\it Class.Quant.Grav.} {\bf 25} 154005


\bibitem{PKVK} Page D N, Kubiz\v n\'ak D, Vasudevan M, and Krtou\v{s}
P 2007 {\it Phys. Rev. Lett.} {\bf 98} 061102



\bibitem{KKPV} Krtou\v{s} P, Kubiz\v n\'ak D, Page D N, and Vasudevan
M 2007 {\it Phys. Rev.} D {\bf 76} 084034

\bibitem{o1}
Houri T, Oota T and Yasui Y 2008
{\it J. Phys.} A  {\bf 41} 025204


\bibitem{HOY:09} Houri T, Takeshi Oota T and Yasui Y 2008 {\it
Phys.Lett.} B {\bf 666} 391


\bibitem{o2} Houri T, Oota T and Yasui Y 2007
{\it Phys. Lett.}  B {\bf 656} 214 



\bibitem{hoi1} Houri T, Oota T and Yasui Y, 2009 {\it
Class.Quant.Grav.} {\bf 26} 045015


\bibitem{KFK} Krtous P, Frolov V P,  and Kubiznak D 2008
{\it Phys.Rev.} D {\bf 78} 064022




\bibitem{FKK} Frolov V P, Krtou\v{s} P, and Kubiz\v n\'ak D 2007
{\it J. High Energy Phys.} {\bf 02} ) 005


\bibitem{OY} Oota T and Yasui Y 2008 {\it Phys.Lett. } B {\bf 659} 688


\bibitem{SKF} Kubiznak D and Frolov V P 2008 {\it JHEP} {\bf 0802} 007




\bibitem{oy2} Oota T and Yasui Y 2008 {\it Int.J.Mod.Phys.} A {\bf 25} 3055


\bibitem{CFK} Connell P, Frolov V P and Kubiznak D 2008 {\it
Phys.Rev.} D {\bf 78} 024042

\bibitem{KFC} Kubiznak D, Frolov V P,  Krtou\v{s} P, and  Connell P  2009
{\it Phys.Rev.} D {\bf 79} 024018
 

\bibitem{FFKK} Frolov V and Krtou\v{s} P, {\it Charged particle in
higher dimensional weakly charged rotating black hole spacetime} 2010
{\it arXiv:1010.2266 [hep-th]}

\bibitem{HHOY} Hamamoto N, Houri T, Oota T, and Yasui Y  2007 {\it J.
Phys.} A {\bf 40},  F177 

\bibitem{KKY:09} Kubiznak D, Kunduri H K and Yasui Y 2009 {\it
Phys.Lett.} B {\bf 678} 240

\bibitem{CCLP} Chong Z W, Cvetic M, Lu H and Pope C N 2005 {\it Phys.
Rev. Lett.} {\bf 95} 161301



\end{thebibliography}
\end{document}